\documentclass[preprint,showpacs,showkeys,preprintnumbers,amsmath,amssymb]{revtex4}

\usepackage{graphicx}
\usepackage{dcolumn}
\usepackage{natbib}
\usepackage{color,calc}
\usepackage{bm}
 
\begin{document}

\title{Pressure-induced order-disorder phase transition in superconducting CaC$_6$}
\author{A. Gauzzi$^1$}
\email{andrea.gauzzi@upmc.fr}
\author{N. Bendiab$^1$}
\author{M. d\textquoteright Astuto$^1$}
\author{B. Canny$^1$}
\author{M. Calandra$^1$}
\author{F. Mauri$^1$}
\author{G. Loupias$^1$}
\author{N. Emery$^2$}
\author{C. H\'{e}rold$^2$}
\author{P. Lagrange$^2$}
\author{M. Hanfland$^3$}
\author{M. Mezouar$^3$}
\affiliation{$^1$Institut de Min\'{e}ralogie et de Physique des Milieux Condens\'{e}s, Universit\'{e} Pierre et Marie Curie-UMR 7590 CNRS, Paris, France}
\affiliation{$^2$Laboratoire de Chimie du Solide Min\'{e}ral-UMR 7555 CNRS, Universit\'{e} Henri Poincar\'{e} Nancy I, 54506 Vandoeuvre-l\`{e}s-Nancy, France\\}
\affiliation{$^3$European Synchrotron Radiation Facility, BP 220, 
38043 Grenoble, France}

\date{\today}
\begin{abstract}
By means of synchrotron X-ray diffraction, we studied the effect of high pressure, $P$, up to 13 GPa on the room temperature crystal structure of superconducting CaC$_6$. In this $P$ range, no change of the pristine space group symmetry, \textit{R\=3m}, is found. However, at 9 GPa, i.e. close to the critical value at which a large $T_c$ reduction was reported recently, we observed a compressibility jump concomitant to a large broadening of Bragg peaks. The reversibility of both effects upon depressurization and symmetry arguments give evidence of an order-disorder phase transition of second order, presumably associated with the Ca sublattice, which provides a full account for the above $T_c$ reduction. 
\end{abstract}


\maketitle
The observation of superconductivity with remarkably high critical temperature $T_c$=11.5 K in CaC$_6$ \cite{wel,eme1} renewed the interest in graphite and related compounds as a model system with a flexible layered structure and tunable electronic and transport properties. Penetration depth \cite{lam}, tunneling \cite{ber} and specific heat \cite{kim3} experiments and Density Functional Theory (DFT) calculations \cite{maz,cal} consistently point at a conventional $s$-wave BCS scenario with a sizable electron-phonon coupling constant, $\lambda\approx 0.8$, that mainly involves two phonon modes with in-plane Ca and out-of-plane C polarizations. According to further studies \cite{cal2,kim2} on the whole series of $A$C$_6$ compounds hitherto synthesized containing alkaline earths $A$=Sr,Ca,Ba \cite{gue,eme2,kim2}, the largest $T_c$ found in CaC$_6$ arises from a stronger electron-phonon coupling between the interlayer band and the above phonons. In addition, the above studies suggest a scaling of $T_c$ with the inverse of the interlayer distance that can be tuned either by pressure or by the $A$ cation size.     

Consistently with this scenario, $T_c$ is found to increase with pressure in all $A$C$_6$ compounds, i.e. CaC$_6$ \cite{gau,smi,kim}, SrC$_6$ \cite{kim2} and YbC$_6$ \cite{akr}. CaC$_6$ exhibits the largest increase up to $T_c$=15.1 K, the highest value for graphite-related compounds, at 7.5 GPa \cite{gau}. At $P\approx 8$ GPa, a sudden $T_c$ drop down to $\approx$5 K concomitant to an anomaly of the aforementioned Ca and C phonons suggests a picture of pressure-induced structural instability limiting $T_c$ \cite{gau}. This picture is supported by the theoretical prediction of a vanishing mode frequency for the in-plane Ca phonon under high pressure \cite{cal2,kim}. A similar instability seems to occur also in YbC$_6$ at $P=2$ GPa \cite{akr}. Whilst the above DFT calculations provide a consistent description of the electronic structure, controversial predictions were made as to the $P$-dependence of the phonon modes and of $T_c$ \cite{cal2,zha,kim2,boe}. For instance, the $P$-dependent $\lambda$ values are systematically underestimated considering the large increase of $T_c$ found experimentally. Also, the structural instability responsible for the drastic $T_c$ reduction at 8 GPa and its effect on the relevant BCS parameters remain to be determined. A possible reason of these difficulties is the inaccurate calculation of the high-pressure structure that remains to be determined experimentally.

In order to address this point, in this work we carried out a detailed study of the room temperature crystal structure of CaC$_6$ at high pressures up to 13 GPa. At 9 GPa, i.e. close to the borderline between the high- and low-$T_c$ phases, we found a jump of the compressibility concomitant to a large Bragg peak broadening. We show that these results put into evidence an order-disorder phase transition of second order that accounts for the above $T_c$ reduction and structural instability at 8 GPa. To our knowledge, this is the first case of superconductor-superconductor phase transition driven by an order-disorder structural transition.    

The experiment was carried out at the high-pressure diffraction beamline ID9 of the ESRF. A high-quality $\approx$1 mm size bulk sample of CaC$_6$ was prepared from a platelet of highly oriented pyrolithic graphite, as described elsewhere \cite{eme2}. A few $\approx$50-100 $\mu$m pieces were mounted in two diamond anvil cells (DAC) in the opposite anvil configuration. The two DACs were subsequently charged with high purity helium (sample n.1) and argon (sample n.2), used as pressure-transmitting media. Due to the reactivity of CaC$_6$, the procedure was carried out in a high-purity dry box, where the cells were sealed under an initial gas pressure of about 0.1 GPa.

The diffractograms were taken in the Debye-Scherrer geometry suitable for the above DAC configuration using a wavelength $\lambda=0.4133$ \AA. The diffracted signal was collected within a maximum diffraction angle 2$\vartheta_{max}=25^{\circ}$ by an image plate. Owing to the small diameter of the stainless steel gasket (150 $\mu$m), the beam spot size was reduced down to 40 $\mu$m for $P\leq$ 10 GPa. At higher $P$, this size was further reduced to 20 $\mu$m due to the shrinkage of the cell. For each diffractogram, the $P$ value was determined from the emission line of a ruby crystal. To maximize the number of reflections detected, the DAC was spindled within an angle of $\pm3^{\circ}$ with respect to the incident beam direction. This is especially important in our case, as the layered structure of CaC$_6$ is expected to induce a preferential orientation of the graphene layers perpendicular to the incident beam. For both samples, no significant amount of secondary phases was detected in the sample regions probed by the X-ray beam. This enabled us to reliably refine the unit cell up to 13 GPa. Sample n.1 gave a larger signal-to-background ratio, so the analysis below mainly refers to this sample. More than hundred diffractograms were taken on sample n.1 by progressively increasing $P$ from 0.1 up to 13 GPa and upon depressurization. Representative diffractograms are shown in Fig. 1. All of the main Bragg peaks are found to match those of the rhombohedric \textit{R\=3m} phase \cite{eme2}. The indexation reported in Fig. 1 corresponds to the equivalent hexagonal unit cell with $a$=4.330 \AA~ and $c$=13.572 \AA~ \cite{eme2}. This cell choice enables to better visualize the crystal structure as an alternate stacking of graphene and Ca layers (see Fig. 1). The additional features appearing above 9 GPa are discussed below.

\begin{figure}
\includegraphics[width=8cm]{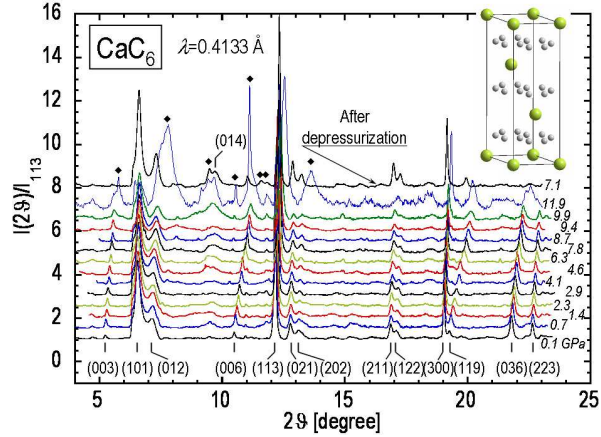}
\caption{(Color online). Representative diffractograms of sample n.1 at different pressures during and after depressurization. Intensities are normalized to the integrated intensity of the strongest (113) peak. Bragg indices refer to the hexagonal unit cell in the inset. Diamonds mark unindexed peaks and additional broad features appearing above 9 GPa (see text).}
\end{figure}

\begin{figure}
\includegraphics[width=8cm]{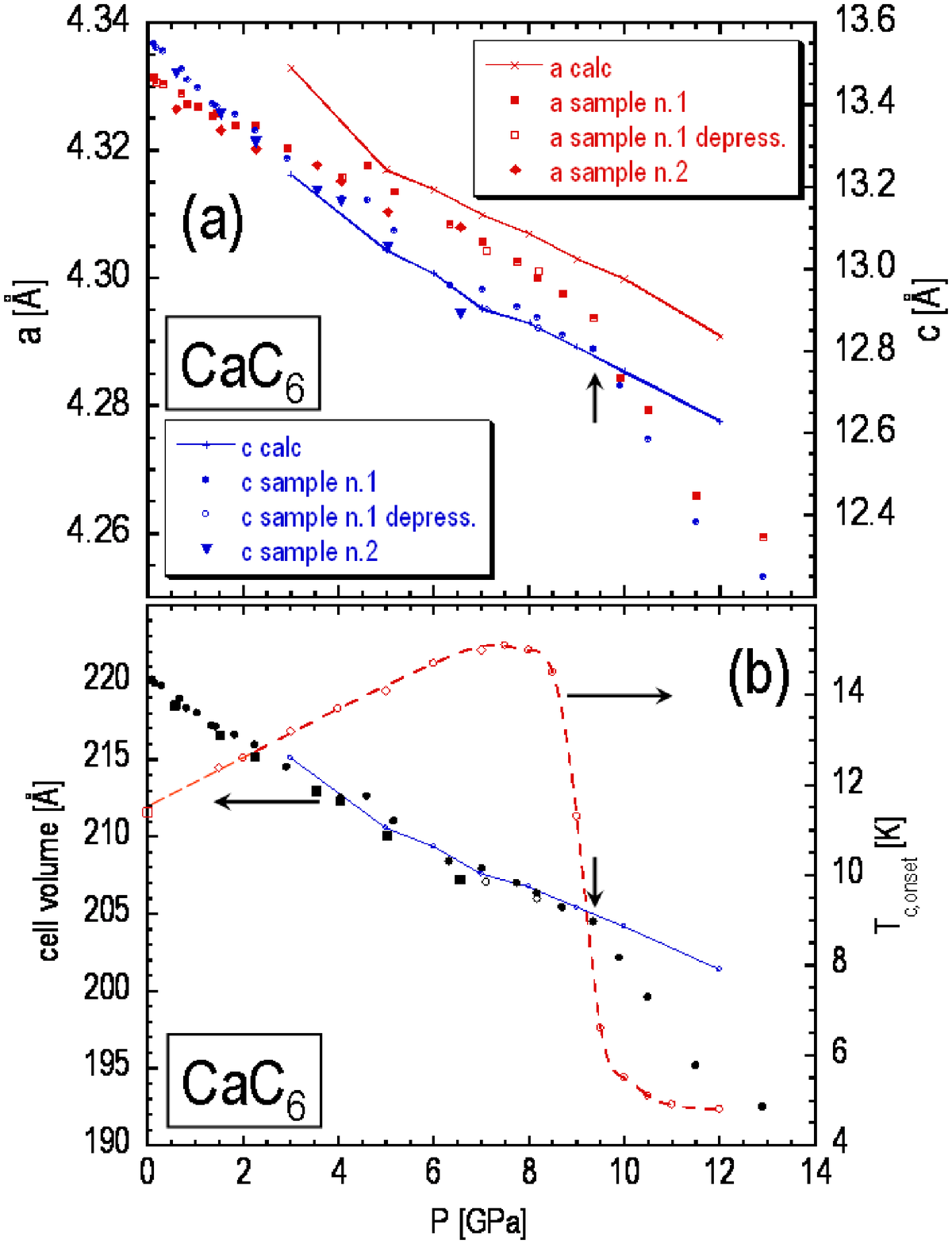}
\caption{(Color online). (a) Experimental (symbols) and calculated (solid lines) pressure-dependence of the hexagonal $a$ and $c$ unit cell parameters. (b) Pressure-dependence of the unit cell volume corresponding to the data (symbols) and calculations (solid line) of panel (a). The $T_c$ vs. $P$ data from ref. \cite{gau} are also reported to show the coincidence of the midpoint of the $T_c$ drop with the kink of the volume compressibility at 9 GPa marked by vertical arrows. The broken line is a guide to the eye.}
\end{figure}

The indexation of Fig. 1 shows that, despite the expected preferential orientation of the sample, reflections with various combinations of $h$, $k$ and $l$ indices are present. This might be due to the folding of the CaC$_6$ flakes during sample manipulation. About thirty diffractograms in the 0.1-13 GPa range were selected for the unit cell refinement. For the $P\lesssim 10$ GPa data, we typically used as many as 12-15 Bragg reflections. This number reduces to 8 at higher pressures because of the weaker signal and the appearance of additional broad features to be discussed below. The result of the refinement and the resulting equation of state are shown in Fig. 2. Thanks to the presence of strong peaks, e.g. (300), (113), (003) and (006), corresponding to independent Bragg planes, the accuracy obtained for the $a$ and $c$ parameters is as good as $\pm 0.001-0.002$ and $\pm 0.002-0.004$ \AA, respectively. Good refinements were obtained also using the data of samples n. 1 and n. 2 upon depressurization and pressurization, respectively. By comparing these three data sets, one notes that the data are reproducible within the experimental error and no hysteresis is found after depressurization.    

Fig. 2 indicates that both $a$ and $c$ parameters decrease monotonically and roughly linearly with $P$ down to 9 GPa. In this $P$-range, a linear regression yields the following values of isothermal compressibility coefficients: $da/dP$=-0.0038 \AA/GPa; $dc/dP$=-0.081 \AA/GPa. Hence, the $c$-axis compressibility is about twenty times larger than the in-plane one, as expected considering the layered structure. The corresponding volume compressibility is $\kappa=(1/V_{\rm 9GPa})dV/dP=-0.0082$ GPa$^{-1}$. The above values and the large anisotropy are consistent with previous pressure studies on graphite \cite{dre} and graphite intercalated compounds with same composition and similar crystal structure, such as LiC$_6$ \cite{cla}. In order to study in detail the $P$-induced structural changes, the relaxed structure was calculated as a function of pressure up to 12 GPa by means of the ESPRESSO implementation \cite{esp} of the DFT, as described elsewhere \cite{cal,cal2}. The results of the calculations are compared with the experimental data in Fig. 2. For $P\lesssim$ 9 GPa, it is noted a good agreement for both $a$ and $c$ parameters, except a slight overestimation of the former.

Most notable feature of the experimental equation of state of Fig. 2 is a kink at 9 GPa corresponding to a jump of the isothermal compressibility, $\Delta\kappa$, the signature of a second order phase transition. The kink leads to a discrepancy between experimental and calculated data in the $P\gtrsim$ 9 GPa range, for, in this range, we performed only volume optimization calculations which do not take into account possible structural instabilities. A linear fit of the experimental data in the $P\gtrsim9$ GPa region yields $\kappa=(1/V_{\rm 9GPa})dV/dP=-0.0215$ GPa$^{-1}$, i.e. almost 3 times larger than in the $P\lesssim9$ GPa region. In order to interpret this result, we notice that no change of space group symmetry is seen in the diffractograms of Fig. 1. Thus, the jump of $\kappa$ at 9 GPa suggests a structural change allowed by symmetry leading to a discontinuity of the electronic structure. It turns out that the only free parameter of the \textit{R\=3m} structure is the in-plane $x$ coordinate of the C atoms in the hexagonal cell \cite{tables}. Thanks to this simple symmetry argument and considering the rigidity of the C honeycomb layers within the plane, one concludes that the only atomic displacement allowed by symmetry is the off-centering of the Ca atoms with respect to the C layers. At ambient pressure, the Ca atoms are centered with respect to these layers \cite{cal2}, as expected from a simple consideration of energy minimization. This implies $x=1/3$, whilst the Ca position is (0,0,0). However, an off-centered in-plane displacement of the Ca atoms requires a modest amount of energy due to the soft phonon mode associated with this displacement \cite{cal}. This scenario is indeed supported by a recent DFT study by Csanyi \textit{et al.} on the structural stability of CaC$_6$ at high pressure \cite{csa}.      

According to the above picture, the large broadening of all peaks concomitant to the kink at 9 GPa (see Fig. 3) shows that the in-plane displacements of the Ca atoms are highly disordered. The broadening turns out to be intrinsic, as it is completely reversible upon depressurization. We conclude that, at 9 GPa, an order-disorder transition of the Ca sublattice occurs. A simple empirical account for this transition is as follows. The $P$-induced compression of the unit cell along the $c$-axis leads to the reduction of the interlayer volume available for the Ca atoms. Because of the relatively large size of the Ca ions, this volume reduction further reduces the limited stability of the potential minima associated with the symmetric positions. Beyond a certain point, any position between adjacent graphene layers becomes energetically equivalent. This disordering mechanism is consistent with the previous observation of anomalous hardening of the out-of-plane C mode and of bad metallic properties with high residual resistivities and flat resistivity curves above 8 GPa \cite{gau}.

\begin{figure}
\includegraphics[width=8cm]{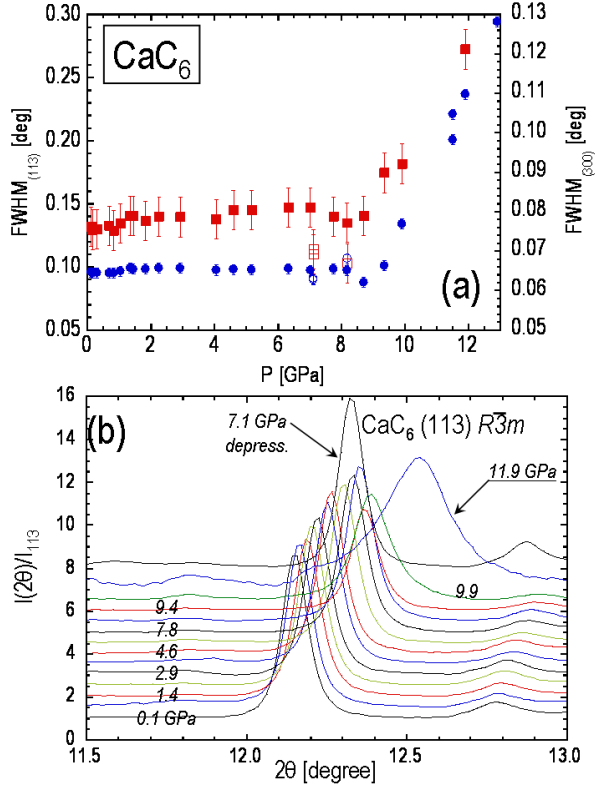}
\caption{(Color online). (a) Pressure-dependence of the full width at half maximum (FWHM) of the (113) (circles) and (300) (squares) peaks. Open symbols refer to data after depressurization. Error bars are smaller than symbols when not visible. (b) Close-up view of the evolution of the normalized integrated intensity of the (113) peak upon pressure. Note the recovery of the pristine width and maximum intensity after depressurization.}
\end{figure}

In addition to the off-centering of the Ca atoms, the study by Csanyi \textit{et al.} \cite{csa} also predicts a $P$-induced buckling of the C layers, thus leading to a lowering of the crystal symmetry. Our data do not enable to verify this prediction, for this would require detecting weak extra peaks, probably below the threshold of our diffractograms. Though, an important insight into the high-$P$ structure is gained by analyzing the new features appearing above 9 GPa. These features must be intrinsic, for they disappear upon depressurization. Most of them are extremely broad, which gives evidence of an incipient amorphous phase. The only changes upon depressurization concern the relative intensity of some peaks. Specifically, the doublet at $2\vartheta \approx 9.5^{\circ}$ (the higher angle peak of the doublet being the (104) reflection) becomes stronger, whilst the (003), (036) and (223) peaks almost vanish. The following two phenomena may account for these changes: 1. a $P$-induced texture change, which typically is irreversible; 2. a partial transformation of the pristine \textit{R\=3m} phase into the competing hexagonal $P6_3/mmc$ one, which would require a modest energy difference \cite{cal2}. Indeed, the diffraction patterns of the two phases are almost identical, but the peak intensities are different. For example, the (002) reflection of the $P6_3/mmc$ phase, which corresponds to the (003) one of the \textit{R\=3m} phase, is very weak, in agreement with our observation. High resolution data on single crystals would help to verify the above scenarios. Finally, our results rule out the scenario of staging transition, which is known to occur in other graphite intercalated compounds at $P\sim$ 0.1 GPa \cite{fis}.

In conclusion, the present high pressure study shows no change of space group symmetry in CaC$_6$ up to 13 GPa. The experimental $P$-dependence of the unit cell parameters is in good agreement with DFT calculations. At 9 GPa, we found a jump of the isothermal compressibility accompanied by a large Bragg peak broadening. These two results give evidence of a $P$-induced order-disorder phase transition. Space group symmetry considerations supported by \textit{ab initio} calculations suggest that the transition should involve the Ca sublattice. This scenario provides a full account of the large $T_c$ reduction and of the large increase of residual resistivity reported previously. Finally, the present study unveils a rather unique type of superconductor-superconductor transition driven by an order-disorder structural transition.       
         
\begin{acknowledgments}
We gratefully acknowledge the ESRF for support (experiment n. HS3120) and thank M. Gauthier, Y. Le Godec, A. Polian and A. Shukla for their valuable comments and suggestions and Y. Grin, K.A. Mueller, U. Schwarz, and H. Takagi for stimulating discussions. 
\end{acknowledgments}

\end{document}